# Timelines for In-Code Discovery of Zero-Day Vulnerabilities and Supply-Chain Attacks

ANDREW J. LOHN, RAND Corporation

Zero-day vulnerabilities can be accidentally or maliciously placed in code and can remain in place for years. In this study, we address an aspect of their longevity by considering the likelihood that they will be discovered in the code across versions. We approximate well-disguised vulnerabilities as only being discoverable if the relevant lines of code are explicitly examined, and obvious vulnerabilities as being discoverable if any part of the relevant file is examined. We analyze the version-to-version changes in three types of open source software (Mozilla Firefox, GNU/Linus, and glibc) to understand the rate at which the various pieces of code are amended and find that much of the revision behavior can be captured with a simple intuitive model. We use that model and the data from over a billion unique lines of code in 87 different versions of software to specify the bounds for in-code discoverability of vulnerabilities - from expertly hidden to obviously observable.

**KEYWORDS**

Supply-Chain, Zero-Day, Vulnerability, Longevity, Discovery, Digital Security,

## 1 INTRODUCTION

In cyber security, the most difficult attacks to defend are those that exploit vulnerabilities in the code that are only known to the attacker (called zero-days). In these cases, there is little that can be done defensively, and even understanding the degree of risk and exposure to zero-days [1] is notoriously difficult because, by their nature, much about them is unknown to anyone other than the attacker. The most frequently considered zero-day vulnerabilities are accidental bugs in the code that are eventually discovered by a potential attacker, but that is only one creation mechanism for zero-days and not the most nefarious one. A growing and concerning trend is for these vulnerabilities or malicious pieces of code to be injected into the legitimate code-base to be distributed to unsuspecting consumers. [2] That could be achieved by an insider or by a targeted cyberattack, and the injected vulnerability may be very easy to identify or it could be very subtle. In August of 2017, Kaspersky researchers found a malicious module embedded in legitimate server software supplied by NetSarang that was used to exfiltrate user data. [3] In September of 2017, Morphisec discovered a malicious modification that created a backdoor in Avast's CCleaner application, turning it into a reconnaissance tool and a dropper. [4,5] And in late 2015, Juniper disclosed [6] both an obvious and a subtle malicious code incorporation that were discovered in the ScreenOS operating system used for firewalls in their routers.

Of the two ScreenOS vulnerabilities, one was a simple backdoor that provided a hardcoded password which would have been easy to discover by even a casual reading of the code. The other was a very subtle change to a section of the encryption algorithm which enabled leakage of information that would help in a decryption effort. Neither of these vulnerabilities existed in earlier versions but once introduced they remained in the code base for years. [6] The attacks highlighted above occurred in proprietary and closed source software, but the opportunity to carry out these attacks may be even higher for open source software.

Anyone can contribute to an open source project so the insider threat is very high and malicious code incorporation risk is therefore very high as well. There have been several recent instances of malicious code being injected into Python repositories and subsequently being incorporated in the code that uses those libraries. [7] Given the popularity of open source software and its extensive incorporation, either in part or in whole, in closed source software it is reasonable to expect a rise in malicious code incorporations targeting open source software as well. These attacks could be extremely far reaching if they target pieces of the fundamental software infrastructure as evidenced by the huge impact of the Shellshock [8] vulnerability in the GNU/Linux operating system, Heartbleed [9] in openSSL, and the bug commonly referred to as simply "the glibc vulnerability," [10] referencing the open source programming library where it is found.

Given the impact that these vulnerabilities can have and that so little can be done about them, an important question is how long is the exposure period once the zero-day exists. For traditional zero-days the attacker must discover the vulnerability and create an exploit for it before the timeline starts, whereas for malicious code incorporation it is reasonable to believe that an exploit was already prepared at the time of the vulnerability's introduction. In both cases the timeline ends once the vulnerability is removed by the defender which can happen in one of four different ways. (1) The attacker can be caught exploiting the vulnerability, (2) the vulnerability can be discovered by defensive testing, (3) the vulnerability can be discovered during code revision, or (4) the vulnerability can be removed without discovery by incidental code revision. The majority of research to date has been dedicated to understanding the first two methods [11,12] but in this paper we focus on the latter two relating to code revision.

To address those questions, we explore the archival code base of several prominent open source projects of different types. We consider an application (Mozilla Firefox), an operating system (GNU Linux), and a programming library (glibc) as summarized in Table 1. The ability to discover any vulnerability during code review should depend on details about the obviousness of the malicious code and the person doing the review, but insight about fundamental trends can be gathered by considering the extreme behaviors. For our analysis, we treat cleverly disguised attacks as only being discoverable if a line of code containing the attack is altered in some way. Simpler attacks such as the introduction of a hardcoded password we treat as being discoverable if any part of the file containing the attack code is altered. The majority of attacks or vulnerabilities should lie between these two regimes.

| Software | Range of Versions | First Date | Last Date | Number of Versions | File Types Examined |
|---|---|---|---|---|---|
| Mozilla Firefox | v1.0 to v47.0 | Nov 2004 | June 2016 | 47 | .cpp, .h, .js |
| GNU | V1.14 to v1.29 | May 2004 | May 2016 | 16 | .c, .h |
| Glibc | V2.0 to v2.23 | Nov 1994 | Feb 2016 | 24 | .c, .h, .sh |

Table 1. Summary of the open source software selected for this study

## 2 METHODOLOGY

The probability of the developers having noticed a subtle attack increases with each additional version release. The metric we use for approximating the discovery of well disguised attacks is based on the lifetime of unique lines of code (uLOC) within the code base. We create a list of all the unique lines of code for each version, throwing out any that are repeated. We then count how many of those uLOC are present in each subsequent version of the software. We use this same metric as an approximation for the probability that a vulnerability is accidentally removed without ever having been discovered, which is essentially the same result as a malicious code incorporation that is so subtle that it is not noticed even as that code is being amended. It may be possible for such a vulnerability to remain even after an edit of the code but we ignore that case in this analysis.

At the other extreme, some attacks may be easy to discover for developers who give even a quick read of the code. To approximate these types of attacks we consider any change to the file containing the zero-day code as sufficient for discovery. We create a list of all the files and their contents and compare them to the same file in the subsequent versions of the software. For each subsequent version, we record the total number of files that have had no changes to their contents or filename while ignoring changes that only affect the filepath and not the file itself.

We use the same fundamental mathematical model to describe both scenarios. We suppose that there could be some fraction of the code that is very unlikely to be examined even over many versions. The rest of the code is examined at a constant rate between all subsequent versions. That process is expressed mathematically in Eq 1 which we use to fit both uLOC and file changes across versions for each file type of each piece of software in Table 1.

$$P = A(1 - e^{-\lambda n}) \qquad (1)$$

The parameter $\lambda$ determines the rate of changes and $A$ is the percent of code that will eventually be changed after the number of subsequent version releases ($n$) becomes sufficiently large. More specifically, the rate parameter ($\lambda$) is the percentage of the changeable code that is altered per version. When comparing the fits for various code bases though, the rate parameters and their effect on the complete code base should be scaled by the percent of code that is potentially changeable ($A$) as can be seen in Eq 2 describing the percent of code altered in any one revision.

$$p = A\lambda e^{-\lambda n} \qquad (2)$$

This relationship comes from taking the derivative of the cumulative percent of code changed (Eq. 1) with respect to the number of subsequent versions to calculate the percent of the code that will be altered in a given version. The prefactor to the exponential provides a very simple means for rough comparison across code bases that is scaled to account for both the rate parameter ($\lambda$) and the saturation level ($A$) as shown in Eq 3.

$$R = A\lambda \qquad (3)$$

We refer to $R$ from Eq 3 as the "base rate" since it is the rate of changes which is then progressively rescaled by the exponential factor from Eq 2 as the code ages. During that ageing process, the probability of removing the zero-day by code revision decreases as the number of subsequent versions increases. That suggests that the longer a zero-day has been in a code base the more likely it is to remain.

This process accurately describes the majority of version changes but we do observe departures from this behavior. In particular, there appears to be a period when a new piece of software is being introduced where newly written code is revised at a higher rate than older code. In some cases, stabilization to a version-independent rate parameter ($\lambda$) occurs within just a few versions and in others it was observed after almost twenty versions. It is advised that practitioners verify whether or not the rate of revision has stabilized as part of the process of extracting parameters for use in modeling.

In addition to the initial stabilization period, there appear to be occassional discontinuous jumps in the rate of change from one version to the next. Those jumps can be significant events in terms of the fraction of code involved, but they are too rare to study statistically with these data. Only 9 such events were observed across all the 424 separate versions of each file type for each of the three types of software in this study which is roughly 2 percent. Estimates made by ignoring those events can provide the baseline trends but should be caveated that rare changes in behavior can occur. In several cases the behavior reverts back to the prior trends following these events, but where there is a clear change in behavior that is maintained consistently afterward we analyze the two patterns separately. After removing these anomalous events from the data, we extract the rate parameter and saturation level using the Nelder-Mead method for maximum likelihood estimation of the log-likelihood.

# 3 RESULTS AND DISCUSSION

## 3.1 Mozilla Firefox

Mozilla Firefox is an open source web browser that was introduced in late 2004 and has been among the most popular browsers for the majority of the last decade. There are over a million unique lines of code in tens of thousands of files as detailed in Fig 1 parts a and b respectively. The largest portion of unique lines of code are written in C++ with javascript being primarily used for aspects related to the user interface events, [13] but the number of javascript files is significantly larger than either of the other two file types.

Such a large code base presents many opportunities for potential malicious code incorporation attacks that may be difficult to find, but the large number of contributors increases the odds of that attack being discovered or otherwise removed. For this paper, we only consider files with extensions .cpp (C++), .js (javascript), or .h (libraries), each of which tend to have distinct roles in the software and may be of varying interest for an attacker and may have different rates of inspection or change to the code.

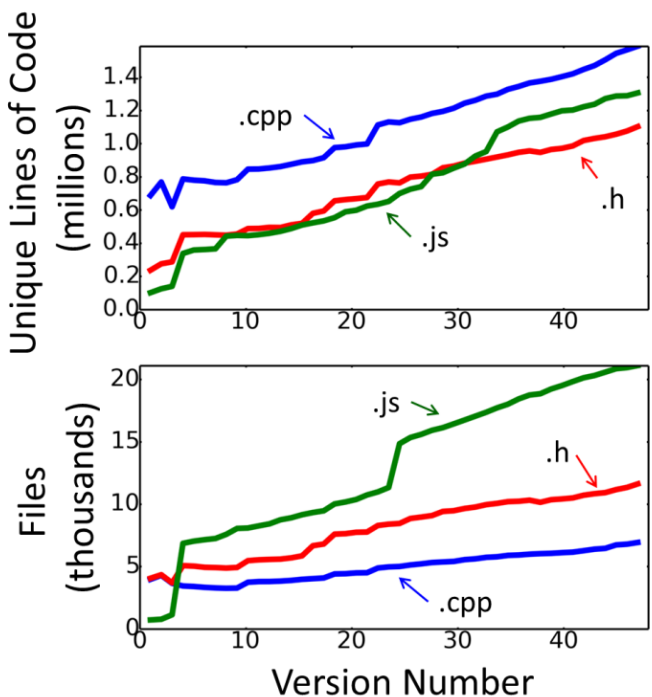

Figure 1. Firefox has been growing continuously both in terms of number of files and number of unique lines of code and is now well over a million unique lines in tens of thousands of separate files.

*3.1.1 C++ Code in Mozilla Firefox.* Figure 2 shows the percentage of .cpp code that was changed from every version to every other version of Firefox. Version 1.0 had 46 subsequent versions up to version 47.0 where each version changed some percentage of the original code. In contrast, version 46.0 had only 1 subsequent version up to version 47.0 so the code was changed by some percent only once. Each curve in the figure depicts the percentage of the original version that has been changed. The older starting versions can be distinguished by having longer curves since there have been more subsequent versions. Additionally, the colors transition progressively from red for the oldest starting versions to blue for newer ones.

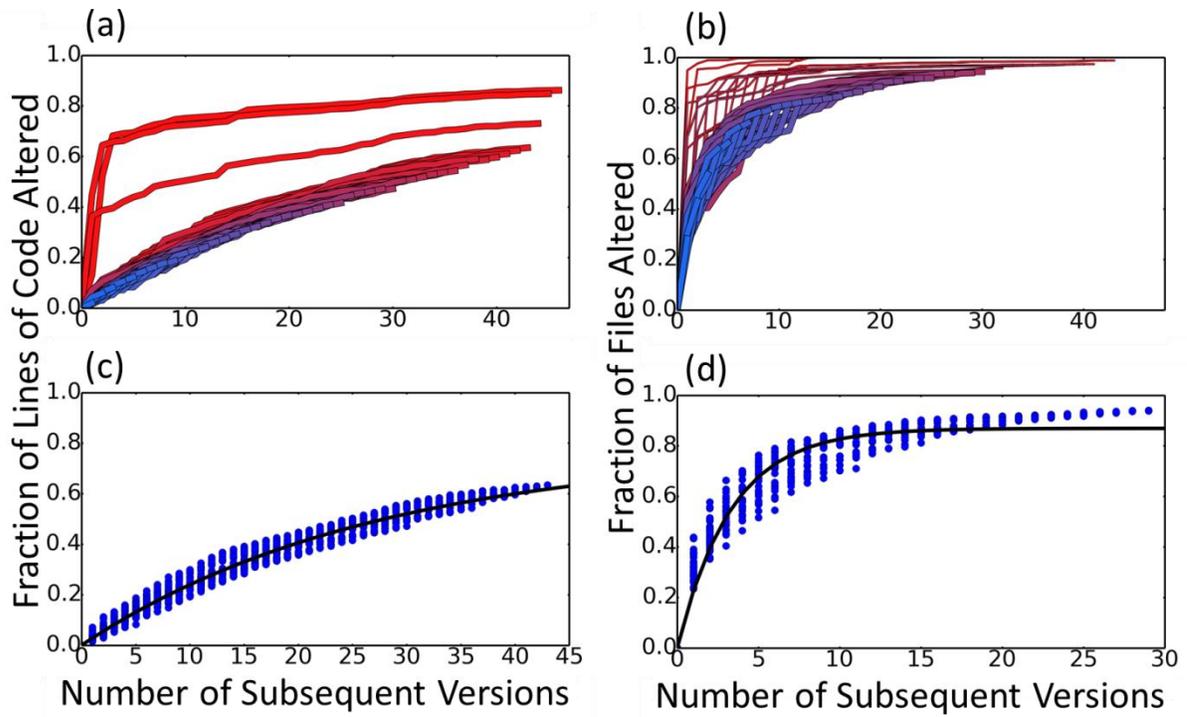

Figure 2. Changes observed from all versions to all subsequent versions for .cpp files in Mozilla Firefox. The earliest versions are shown in red and later versions progressively shift to blue. The percent of changes once code stabilization occurred are plotted and fit in parts c and d along with the fit to Eq 1.

For uLOC shown in Fig 2a, the first two versions have a high initial rate of change and the third version is intermediate before settling on a fairly stable pattern of behavior that lasts until at least the most recent version studied (v47.0). Those first three versions were released at a slower rate than the more contemporary versions with release dates of November 2004, October 2006, and June 2008 respectively. Over the six years and 43 versions since versions since v4.0 in March 2011, about half of all uLOC have been altered. It appears as if the rate of change may be decreasing slightly as the code matures since the bluer curves in Fig 2a are lower than the redder ones, but that trend is not examined in this study to simplify the modeling and avoid overfitting.

The changes to the files shown in Fig 2b are quite different from the behavior observed for uLOC. The first couple versions also have high rates of change but stabilization does not occur until version 18. For visual ease, the curves starting from versions prior to v18 are shown using thinner lines. Even after stabilization, the rate of change is much higher than for uLOC which is not surprising since a change to any line of code constitutes a change in a file.

Fitting the data to Eq 1 enables quantitative comparisons among the various software, file types, and vulnerability types. To focus on the stable trends we remove the versions from before the stabilization and fit the rest as shown in Fig 2c,d. The saturation value for changes uLOC that .cpp files are trending toward is 78% and rate parameter ($\lambda$) for cpp uLOC is 0.0369 giving a base rate of almost 3%. By contrast, the .cpp file changes are trending toward almost 87% changes with a rate parameter of 0.302 and a base rate almost an order of magnitude higher at 26%.

*3.1.2 Javascript Code in Mozilla Firefox.* In some ways the behavior of changes to javascript is qualitatively similar to cpp. The uLOC (Fig 3a) stabilizes after the first three versions but the files remain erratic much longer. After stabilizing, the javascript files (Fig 3b) have two versions with significant jumps, first at version 15 and the second at version 25. To help visually separate the jumps, curves that are affected by the jumps are separated using thin lines from 0 to 3, thicker for 4 to 15, thicker again for 15 to 26 and thickest for later versions. One subtle difference from the .cpp files is that the uLOC changes appear to be trending toward a higher rate instead of a lower one although, again, this aspect of the behavior is not examined further in this study.

For fitting the uLOC, the first three versions were removed to focus on the stable trend in Fig 3c. For fitting the file changes, only data points leading up to one of the jumps or versions that were released after the final jump are included. That process separates the stable part of the trend shown in Fig 3d from the major but rare events that separate the curves in Fig 3b.

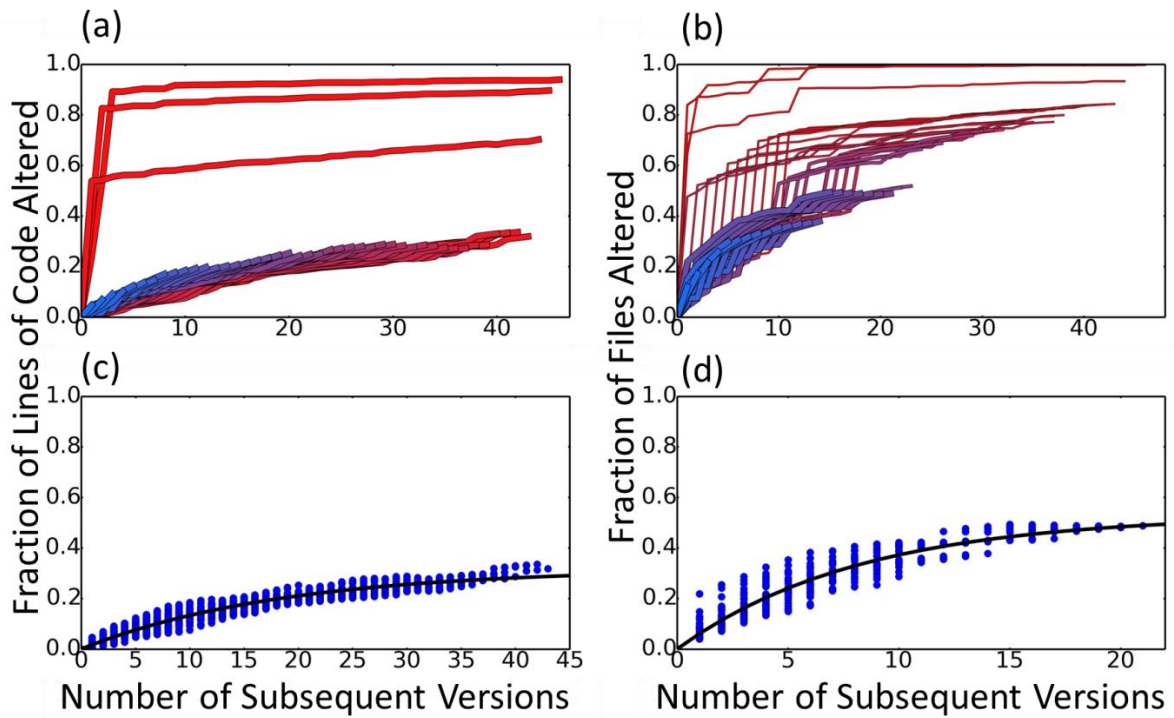

Figure 3. Changes observed from all versions to all subsequent versions for .js files in Mozilla Firefox. The earliest versions are shown in red and later versions progressively shift to blue. The percent of changes once code stabilization occurred are plotted and fit in parts c and d along with the fit to Eq 1.

Over the full range of versions studied, the javascript uLOC reach about 30% as compared to the 60% for the .cpp extensions and the javascript trend appears closer to saturation. According to the fit, the javascript uLOC changes trend toward a saturation value of 32% at a rate parameter of 0.05 for a base rate of about 1.7% which is significantly lower than the 3% for .cpp extensions. For the file changes, the saturation value is 53% with a rate parameter of 0.121 and a base rate of about 6% which are also all lower than the .cpp files.

*3.1.3 H Libraries in Mozilla Firefox.* Changes to the library uLOC (Fig 4a) and files (Fig 4b) show qualitatively similar behavior to both the C++ files and the javascript files. The stabilizations follow the same patterns and the observed rates of change are intermediate between the two.

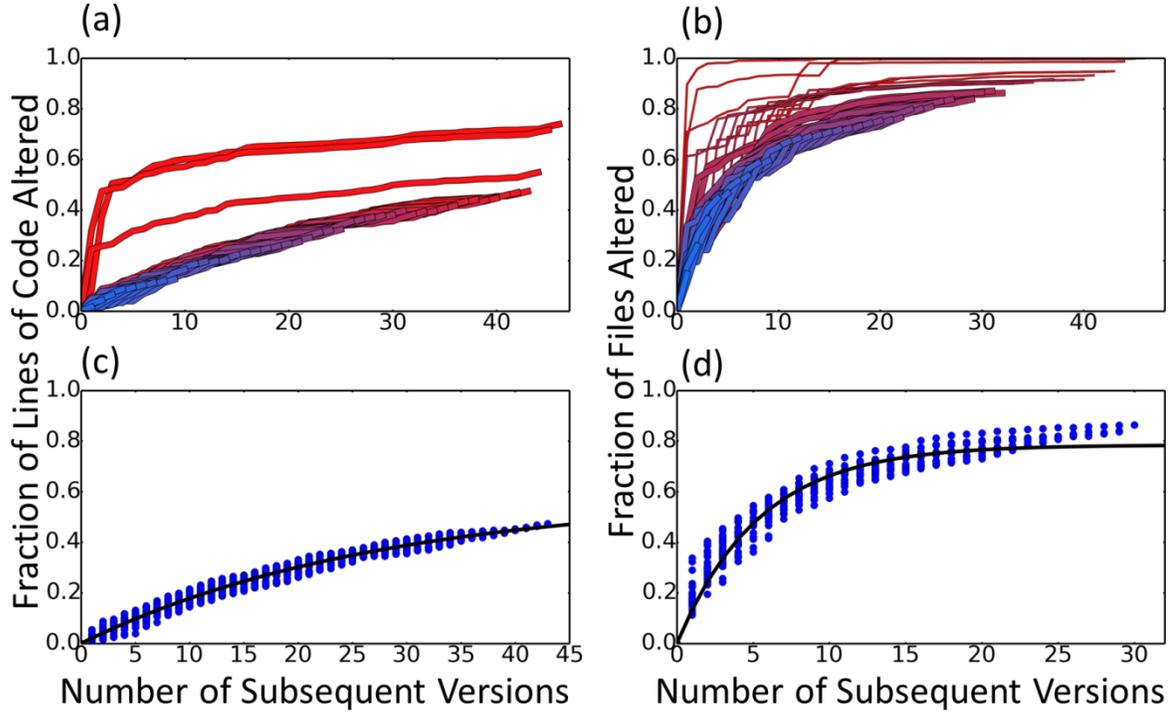

Figure 4. Changes observed from all versions to all subsequent versions for .h files in Mozilla Firefox. The earliest versions are shown in red and later versions progressively shift to blue. The percent of changes once code stabilization occurred are plotted and fit in parts c and d along with the fit to Eq 1.

The same procedure was followed for fitting the .h extensions as was used for the .cpp file extensions where the first three versions were excluded for the uLOC (Fig 4c) analysis but the fit starts at v17 instead of v18 for the file changes (Fig 4d). The uLOC changes trend toward 58% at a rate of 0.036 for a base rate of 2% and the file changes trend toward 78% at a rate of 0.19 for a base rate of 15%. Both the base rates and saturation values are intermediate between the .cpp and .js extensions.

| Code Type | Rate Parameter ($\lambda$) | Saturation Level ($A$) | Base Rate |
|---|---|---|---|
| Cpp uloc | 0.0369 | 0.777 | 0.0287 |
| Cpp Files | 0.302 | 0.869 | 0.262 |
| Js uloc | 0.0541 | 0.318 | 0.0172 |
| Js Files | 0.121 | 0.531 | 0.0643 |
| H uloc | 0.0363 | 0.584 | 0.0212 |
| H Files | 0.186 | 0.784 | 0.146 |

Table 2. Lifetime parameters extracted from the Firefox data

For all three of .cpp, .js, and .h extensions and for both uLOC and changes at the file level, the early versions experienced high levels of change before stabilizing at lower levels. For uLOC the stabilization occurred by version 4, but for the file changes that stabilization occurred much later. Ignoring the early behavior and focusing on the stable trend, 30 to 60 percent of the uLOC were changed at some point in the 44 versions. For the files that number was higher, ranging from around 50 to 90 percent. With the exception of the two distinct jumps that were observed in the javascript file level changes, the behavior can be fairly well captured by the simple dynamics described in Eq 1. Those two jumps are worth note however because they represent significant departures from the trend.

### 3.2 GNU/Linux Operating System

GNU/Linux is an operating system that is freely available and open source. It is built on the Linux kernel and GNU/Linux itself is used in a wide variety of not just personal computers but servers and embedded systems including those on the international space station for example. [14] It was started in September 1983, made to be compatible with Unix, and named GNU for "GNU Not Unix." The versions considered in this study are available in the GNU Project's repository ranging from version 1.14 released in March 2004 to version 1.29 released in March 2016. [15]

GNU is significantly smaller than Mozilla Firefox in terms of both uLOC (Fig 5a) and number of files (Fig 5b). For GNU we only examined files with extensions .c and .h. Tens of thousands of uLOC exist for .c in hundreds of files whereas .h has from a few thousand to around ten thousand uLOC in about 100 files. There are also only 16 GNU releases studied here compared to 47 for Firefox.

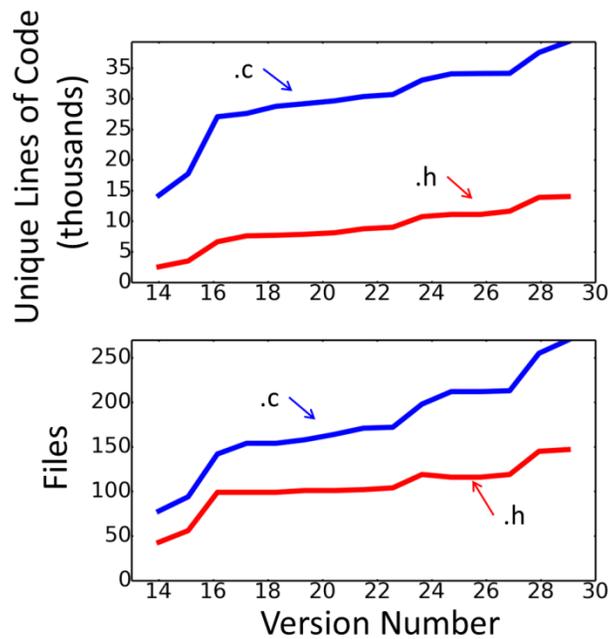

Figure 5. GNU has been growing continuously from versions 1.14 to 1.29 with tens of thousands of unique lines of code in hundreds of files.

*3.2.1 C Code in GNU/Linux.* A consequence of starting at version 1.14 is that if there was a stabilization period as was observed in the early versions of Firefox it would not be observed in this data. The version to version changes for .c uLOC (Fig 6a) and files (Fig 6b) are relatively consistently behaved as a result. Nevertheless, it appears that smaller jumps in the uLOC were taken at versions 1.16 and 1.23 so the equation is fit only to versions up until their first jump (Fig 6c) and all releases following the last jump. Any discrete changes with regard to files were less clear and therefore the all data points are retained for fitting in part d of the Fig 6.

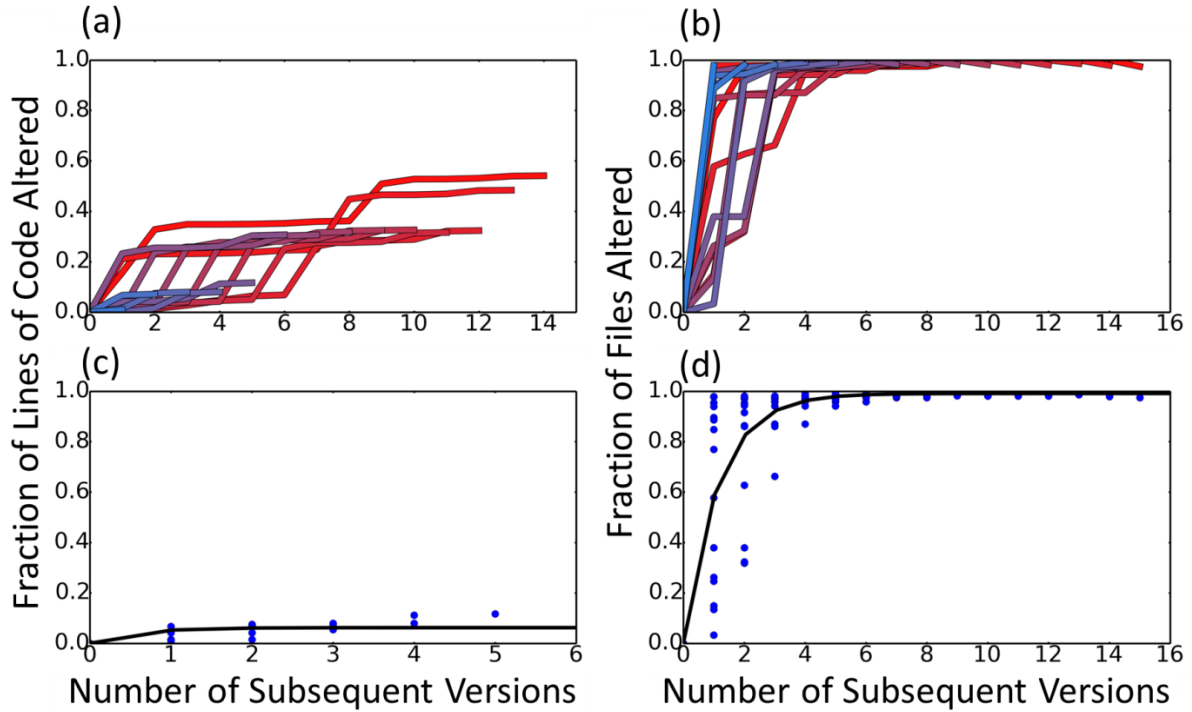

Figure 6. Changes observed from all versions to all subsequent versions for .c files in GNU. The earliest versions are shown in red and later versions progressively shift to blue. The percent of changes once code stabilization occurred are plotted and fit in parts c and d along with the fit to Eq 1.

From version 1.14 to version 1.29 about half of the uLOC were altered at some point. The majority of those changes occurred during the two major events (major code changes can be seen in Fig 6a by the jumps in the first two revisions and in the 9$^{th}$ revision) as opposed to the continuous changes that are modeled here. To predict the lifetime of a malicious code injection over many versions one would need to predict when major overhauls take place but in this case statistical analysis is not possible since there are only two events. It may still be sufficient over shorter periods to consider the more steady-state trend shown in part c, and intuition can be built based on the observation of the relative frequency and effect of those major overhauls.

The steady-state uLOC variation trends toward a saturation of about 6% with a rate parameter of 1.856 for a base rate of around 12%. This high rate parameter and low saturation percentage indicate that few changes are made outside of the rare major changes. On the other hand, the files changed much more rapidly. Nearly all files had been altered in some way within the first few versions. As a result, the saturation was 99% with a rate parameter of 0.883 and a base rate of 88%.

*3.2.3 H Libraries in in GNU/Linux.* The trends in the library (.h) files (Fig 7b) are mostly consistent with those observed in the .c files. Nearly all files were altered in some way within the first few subsequent releases and the uLOC alterations were far less frequent. In terms of uLOC (Fig 7a), the difference between large revisions and steady-state revisions was not as pronounced as they were for the .c files so all data points were retained for the fitting (Fig 7c and 7d).

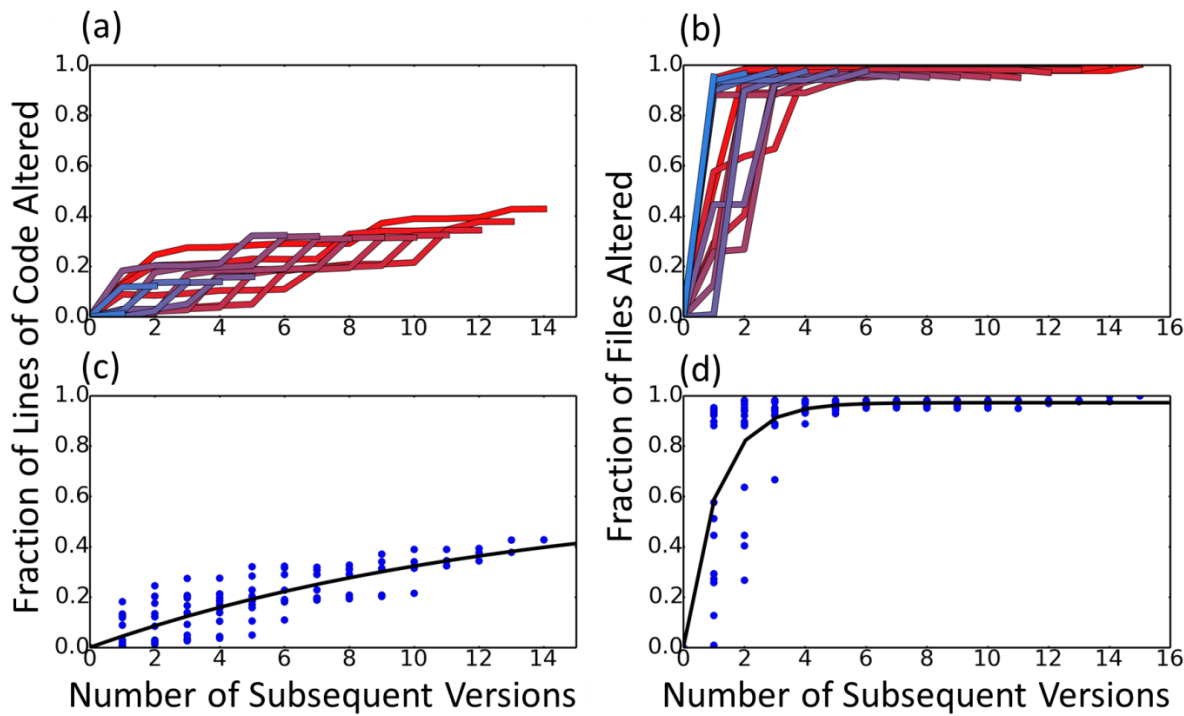

Figure 7. Changes observed from all versions to all subsequent versions for .h files in GNU. The earliest versions are shown in red and later versions progressively shift to blue. The percent of changes once code stabilization occurred are plotted and fit in parts c and d along with the fit to Eq 1.

The uLOC alterations were trending toward a saturation of about 60% with a rate parameter of 0.0757 and a base rate of about 5%. By contrast, the fraction of files amended trends toward 97% with a rate parameter of 0.917 and a base rate of 89%.

The files appear to undergo rapid and continuous revision, but malicious changes made to the GNU code base that would require careful inspection to discover are much more likely to persist across versions if they are in the .c files as opposed to the .h files. The uLOC for .h files is the most interesting as it departs from the extremely low rate of changes in uLOC for .c and the extremely high changes at the file level.

| Code Type | Rate Parameter ($\lambda$) | Saturation Level ($A$) | Base Rate |
|---|---|---|---|
| C uloc | 1.856 | 0.062 | 0.115 |
| C Files | 0.883 | 0.991 | 0.875 |
| H uloc | 0.0757 | 0.608 | 0.046 |
| H Files | 0.917 | 0.972 | 0.891 |

Table 3. Lifetime parameters extracted from the GNU data

Vulnerabilities in GNU that could be identified by someone making any changes to the files are likely to be removed within the first subsequent version or two and almost certain to be removed after a few subsequent versions. Vulnerabilities that are harder to identify, which would only be discovered by changes to the specific lines of code, are much more likely to remain for many versions. That is particularly true for files with the .c extension but also for the .h extensions. Despite having a fairly high saturation level, the changes to uLOC for the .h files average only around two percent per version over the range studied here.

### 3.1 Glibc Programming Library

Glibc is an open source implementation of libraries for the C programming language and has evolved to service other languages as well. It was developed in concert with the GNU operating system and predates the Linux kernel. It has been used in the development of a very wide variety of systems including consumer hardware, routers, programming languages, and embedded devices so vulnerabilities can have a far-reaching and profound impact. As an example, a remotely exploitable buffer overflow bug in glibc that existed in every version starting with v2.9 (Nov, 2008) and was only discovered over seven years later in February 2016. [16] That vulnerability exposed personal computers, routers, and millions of internet-connected devices from a broad range of manufacturers. [10,17]

The historical versions of glibc are freely available [18] and the volume of code is an intermediate between the two other code bases in this study, Mozilla Firefox and GNU. Glibc has hundreds of thousands of uLOC (Fig 8a) in a few thousand files (Fig 8b). In addition to the .c and .h extensions considered for the GNU, we also consider the shell scripts with extension .sh for which the uLOC ranges from 87 to 3027 and the number of files ranges from 2 to 57 with already 33 files by version 2.2. The 24 total versions considered here range from v2.0 released in January 1997 to version 2.23 released in February 2016.

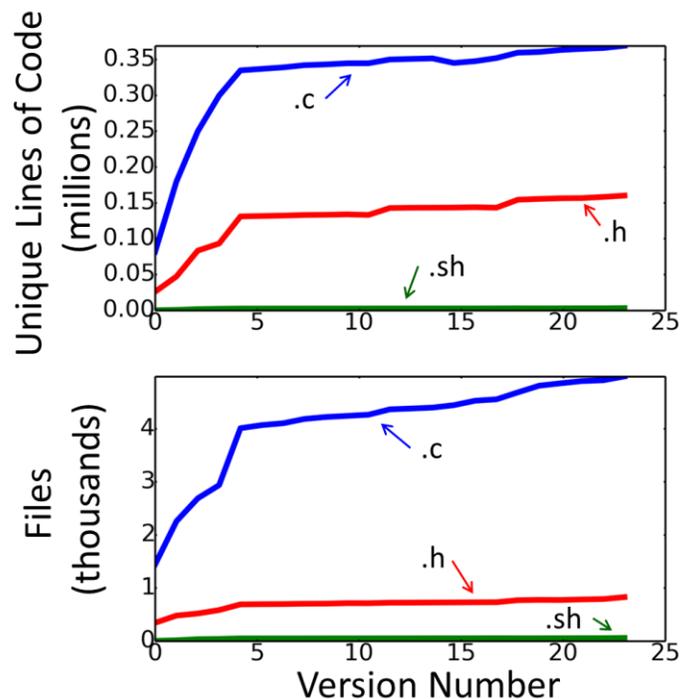

Figure 8. There are hundreds of thousands of lines of code in thousands of files in current versions of glibc as shown for versions 2.0 to 2.23.

*3.3.1 C Code in glibc.* As was observed in the Mozilla Firefox case, for uLOC, the first three versions have significantly different behavior than the later versions which is then fairly consistent (Fig 9a). Unlike Firefox, the file level changes (Fig 9b) reached a steady state much faster. The first three versions had high rates of change with an intermediate fourth version before stabilizing. Also unlike the previous cases glibc file changes reach multiple steady states. The jump observed in Fig 9b in version 2.16 is not just an isolated major event but a distinct change in behavior that carried forward through subsequent versions. In Figure 9 parts c and d we fit the stable state for uLOC and both stable states for the files respectively. The higher rate marked in blue is the more recent trend and the lower rate marked in red was the previous one.

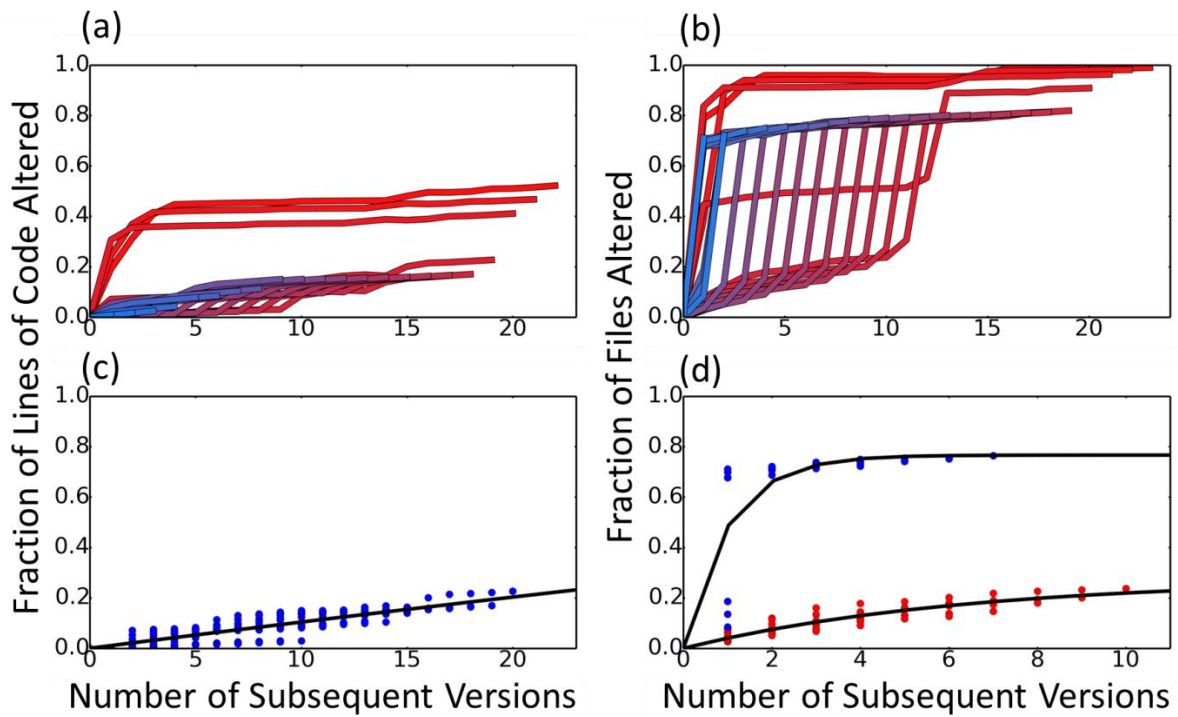

Figure 9. Changes observed from all versions to all subsequent versions for .c files in glibc. The earliest versions are shown in red and later versions progressively shift to blue. The percent of changes once code stabilization occurred are plotted and fit in parts c and d along with the fit to Eq 1.

The first three versions of glibc from this study (v2.0-2.2) were excluded from the fit in Fig 9c to estimate the more recent trend for uLOC. For the remaining versions, the cumulative changes only approach 20%. Additionally, the trend appears more linear than the exponential trends observed for the other software. As a result, the fitting parameters are more difficult to interpret. The saturation parameter is 2.16 and the rate parameter is 0.005 implying a slow trend toward 216% of the uLOC changed, which is impossible. In this case, the base rate which is about 1.1% is a constant rate of change that should be interpreted as the rate of change from one subsequent version to the next. The base rate should not have that interpretation when the trend is nonlinear. The linear trend observed in this case clearly cannot continue forever and the rate must eventually decrease, but it appears to hold over the range of versions examined here.

For the file changes (Fig 9d), in addition to the first three, the fourth version (v2.3) was also left out and two separate trends were fit. The older and lower rate trend was for versions up to v2.15 (shown in red dots) and the more recent high rate of change trend is for versions after v2.15 (shown in blue). The older is trending toward 28% of files changed at a rate of 0.158 for a base rate of 4.4% and the more recent is trending toward a much higher 77% at a rate of 0.99 for a base rate of 76%.

*3.3.2 H Libraries in glibc.* The .h libraries in glibc have very similar behavior to the .c extensions in both uLOC (Fig 10a) and file changes (Fig 10b) with only subtle differences. They have the same initial behavior in the first three versions but the uLOC in the fourth version is intermediate, which was not the case for the .c code.

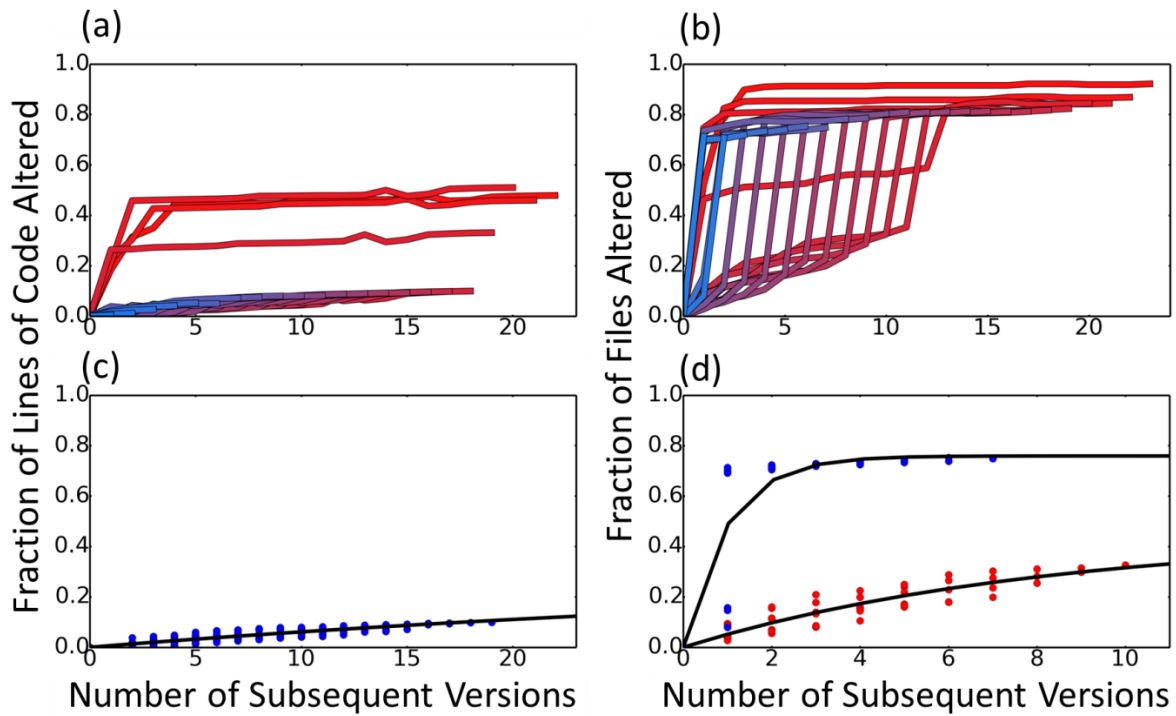

Figure 10. Changes observed from all versions to all subsequent versions for .h files in glibc. The earliest versions are shown in red and later versions progressively shift to blue. The percent of changes once code stabilization occurred are plotted and fit in parts c and d along with the fit to Eq 1.

In fitting uLOC (Fig 10c) for the .h extensions, the first four versions were excluded whereas the fourth version was included in the fitting for the .c extensions. Other than that the procedures were identical. The uLOC data are a little more consistent, which leads to a tighter fit and more reasonable fitting parameters. The uLOC changes are trending toward 31% at a rate of 0.022 for a base rate of 0.7% which is slightly lower than the rate of uLOC changes for the .c code.

For the file changes (Fig 10d), the behavior from versions v2.3 to v2.15 were trending toward 45% at a rate of 0.12 for a base rate of about 5%. The more recent behavior from version v2.16 on is trending toward 76% at a rate of 1.02 for a base rate of 77%. Both of these rates and their longer-term trends are very comparable to those observed for the c files.

*3.3.3 SH Scripts in glibc.* The .sh file extensions for glibc also show behavior that is mostly consistent with both the .c and .h extensions. The uLOC changes (Fig 11a) rapidly reach a stable state that is maintained across all subsequent versions studied, although for .sh files that steady state is achieved earlier, after only two of the observed versions. The file changes (Fig 11b) also behave similarly, with the first three versions undergoing rapid changes and the fourth as an intermediate before reaching a stable state that is maintained until version 2.16.

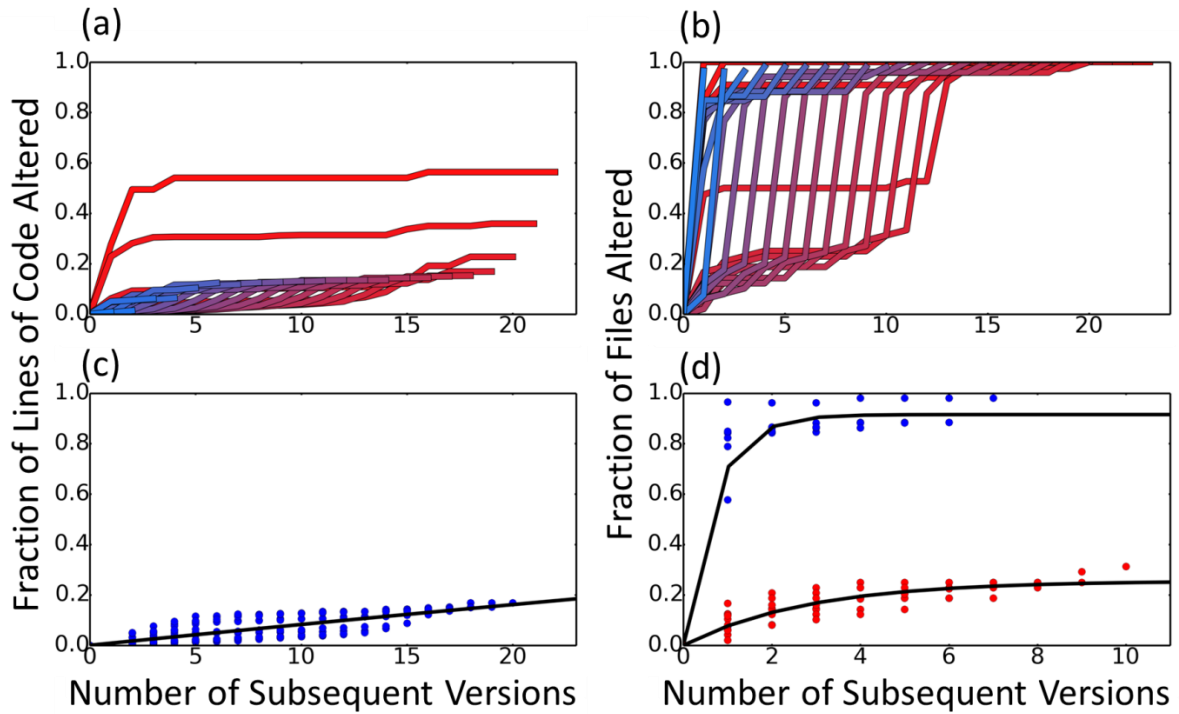

Figure 11. Changes observed from all versions to all subsequent versions for .sh files in glibc. The earliest versions are shown in red and later versions progressively shift to blue. The percent of changes once code stabilization occurred are plotted and fit in parts c and d along with the fit to Eq 1.

The uLOC changes were fit (Fig 11c) to all but the first two versions and the trend is highly linear as was observed for the .c extensions. The changes are trending toward an impossibly high saturation value of 180% at a rate of 0.005 for a base rate of about 0.8% which is close to both the 1.1 and 0.7 observed for .c and .h extensions respectively. As was true for the .c extensions, it is best to treat the trend as linear over the observed range with the base rate as the rate of changes per version.

The files were once again fit (Fig 11d) ignoring the first four versions and treating those up to v2.16 separately from those after. The older behavior trended toward 26% at a rate of 0.357 for a base rate of about 9% and the more recent behavior trends toward 92% at a rate of 1.46 for a base rate of 130%. This 130% is not impossible in the way the linear trends are, it just represents a very large initial rate of change that rapidly decreases before even the first subsequent version is released.

| Code Type | Rate Parameter ($\lambda$) | Saturation Level ($A$) | Base Rate |
|---|---|---|---|
| C uloc | 0.00494 | 2.158 | 0.0107 |
| C Files – High (recent) | 0.991 | 0.766 | 0.759 |
| C Files – Low (old) | 0.158 | 0.276 | 0.0436 |
| H uloc | 0.0216 | 0.315 | 0.0068 |
| H Files – High (recent) | 1.02 | 0.759 | 0.774 |
| H Files – Low (old) | 0.122 | 0.447 | 0.0545 |
| Sh uloc | 0.00472 | 1.791 | 0.00845 |
| Sh Files – High (recent) | 1.46 | 0.915 | 1.33 |
| Sh Files – Low (old) | 0.357 | 0.256 | 0.0914 |

Table 4. Lifetime parameters extracted from the glibc data

Vulnerabilities introduced in glibc are likely to remain for a long time if they require careful inspection or change to a specific line of code because all the uLOC rates are only about one percent per version despite the high saturation levels. The same is not true if the vulnerability can be identified by anyone casually changing the files, although that has only been true for the more recent versions. Prior to version 2.16 changes to the files were also rare and a vulnerability would have been expected to persist for many versions even if it was relatively easy to identify.

## 4 CONCLUSIONS

Through evaluation of the rate of change to over a billion lines of code in over a million files across the history of three different types of open source software, we have assessed the rate of discovery and removal of zero-day vulnerabilities, particularly those that have been maliciously introduced in the code base. Our approach is limited to discovery of the vulnerability by code inspection or to accidental removal of the vulnerability and is not related to detection of the attack in progress or detection through functionality testing.

By treating revision rate as a measure of discoverability we approximate bounds on the probability of in-code vulnerability discovery. The odds of discovery of a vulnerability range from requiring that the specific lines of code containing the vulnerability be amended, to the vulnerability being observable upon any change to the file that contains it. For several of the codes considered here the probability of discovery was reasonably small, at 20% or less even over many version releases. Much of that percent comes in the first few subsequent versions following a new release, after which the code tended to reach a stable rate of change and large fractions of code often remained unedited. The specific numbers for vulnerability lifetime and version-to-version edit rates vary from software to software and even filetype to filetype within the same software, but the stability of the trends indicates that at least the baseline rate of zero-day vulnerability discoverability can be estimated for a piece of software once it has reached a stable state of code revision.

ACKNOWLEDGMENTS

The author would like to thank Lara Schmidt and Joshua Baron for their insight and guidance. This work was supported by the US Air Force under contract FA7014-16-D-1000.